\documentclass[a4paper]{jpconf}
\usepackage{graphicx}
\usepackage[applemac]{inputenc}

\begin{document}

\title{The two-dimensional frustrated Heisenberg model on the 
orthorhombic lattice}

\author{B Schmidt, M Siahatgar, P Thalmeier, and A A Tsirlin}

\address{Max-Planck-Institut für Chemische Physik fester Stoffe, Dresden, Germany}

\ead{bs@cpfs.mpg.de}

\begin{abstract}
We discuss new high-field magnetization data recently obtained by
Tsirlin et al.{} for layered vanadium phosphates in the
framework of the square-lattice model.  Our predictions for the
saturation fields compare exceptionally well to the experimental
findings, and the strong bending of the curves below saturation agrees
very well with the experimental field dependence.  Furthermore we
discuss the remarkably good agreement of the frustrated Heisenberg
model on the {\em square\/} lattice in spite of the fact that the
compounds described with this model actually have a lower
crystallographic symmetry.  We present results from our calculations
on the thermodynamics of the model on the {\em orthorhombic\/} (i.e.,
rectangular) lattice, in particular the temperature dependence of the
magnetic susceptibility.  This analysis also sheds light on the
discussion of magnetic frustration and anisotropy of a class of
iron pnictide parent compounds, where several alternative
suggestions for the magnetic exchange models were proposed.

\end{abstract}

\section{Introduction}

The frustrated $S=1/2$ Heisenberg model on the square lattice, the
$J_{1}$-$J_{2}$ model, appears to describe well the thermodynamic and
magnetic properties of two classes of vanadium compounds of type
Li$_{\rm2}$VO$X$O$_{\rm4}$ ($X =$ Si, Ge)~\cite{melzi:01} and
$AA'$VO(PO$_{\rm4}$)$_{\rm2}$ ($A$, $A' =$ Pb, Zn, Sr,
Ba)~\cite{kaul:04,nath:08}.  They consist of V-oxide
pyramid layers containing V$^{4+}$ ions with $S=1/2$.  From the
analysis of the temperature dependence of the heat capacity and the
magnetic susceptibility in zero (or small) fields, the frustration
ratio $J_{2}/J_{1}$ can be ontained~\cite{shannon:04}.  However, an
ambiguity remains with respect to the relative sign of the two
exchange constants, which can be resolved by analyzing the behavior of
these materials in finite fields~\cite{schmidt:07b}.  The average
exchange constants $J_{\rm c}=\sqrt{J_{1}^{2}+J_{2}^{2}}$ of these
materials are low enough such that their saturation fields are
experimentally accessible.  In this article we discuss the high-field
magnetization of the $J_{1}$-$J_{2}$ model and compare our findings to
recent measurements~\cite{tsirlin:09a}.

Although we describe the physics of the $J_{1}$-$J_{2}$ compounds 
using a square-lattice model, their true crystal structure 
corresponds to a two-dimensional lattice with lower symmetry. 
Therefore, we introduce an additional spatial anisotropy in the ab 
plane assuming
orthorhombic symmetry, i.\,e., 
a rectangular lattice.

\section{Model Hamiltonian}

The effective Hamiltonian on the rectangular lattice has the form
\begin{equation}
    {\cal H} = J_{1a} \sum_{\langle ij \rangle_{1a}} \vec{S}_i\cdot\vec{S}_j
    + J_{1b} \sum_{\langle ij \rangle_{1b}} \vec{S}_i\cdot\vec{S}_j
    + J_2 \sum_{\langle ik \rangle_2} \vec{S}_i\cdot\vec{S}_k
    \label{eqn:ham}
\end{equation}
where $J_{1a}$ and $J_{1b}$ denote the nearest-neighbor exchange along
the $a$ and $b$ directions, and $J_{2}$ labels the diagonal 
next-nearest neighbor exchange. A more convenient parametrization is
\begin{eqnarray}
    J_{1a}&=&\sqrt{2}J_{\rm c}\cos\phi\cos\theta,
    \ 
    J_{2}=J_{\rm c}\sin\phi,
    \nonumber\\
    J_{1b}&=&\sqrt{2}J_{\rm c}\cos\phi\sin\theta,
    \ 
    J_{\rm c}=\sqrt{\frac{1}{2}\left(J_{1a}^2+J_{1b}^{2}\right) + 
    J_2^2},
    \label{eqn:params}
\end{eqnarray}
introducing a frustration angle $\phi$, an anisotropy parameter
$\theta$, and an overall energy scale $J_{\rm c}$.  For
$\theta=\pi/4$, the above Hamiltonian reduces to the square-lattice
case.

The results presented in the following paragraphs are obtained by
applying linear spin-wave theory (LSW) and exact diagonalization (ED)
for finite clusters at finite temperatures
(FTLM)~\cite{shannon:04,schmidt:07b,thalmeier:08}.

\section{Saturation field}

From LSW the saturation field for the square lattice is given by the 
instability of the fully polarized state against a single-magnon 
excitation.
This result is in exact agreement with ED for antiferromagnetic
$J_{2}$.  However, exact diagonalization for our finite clusters
reveals that for ferromagnetic $J_{1}$, a $\Delta S=2$ instability
determines the saturation field~\cite{schmidt:07b}.

\begin{figure}
    \centering
    \includegraphics[height=0.3\columnwidth]{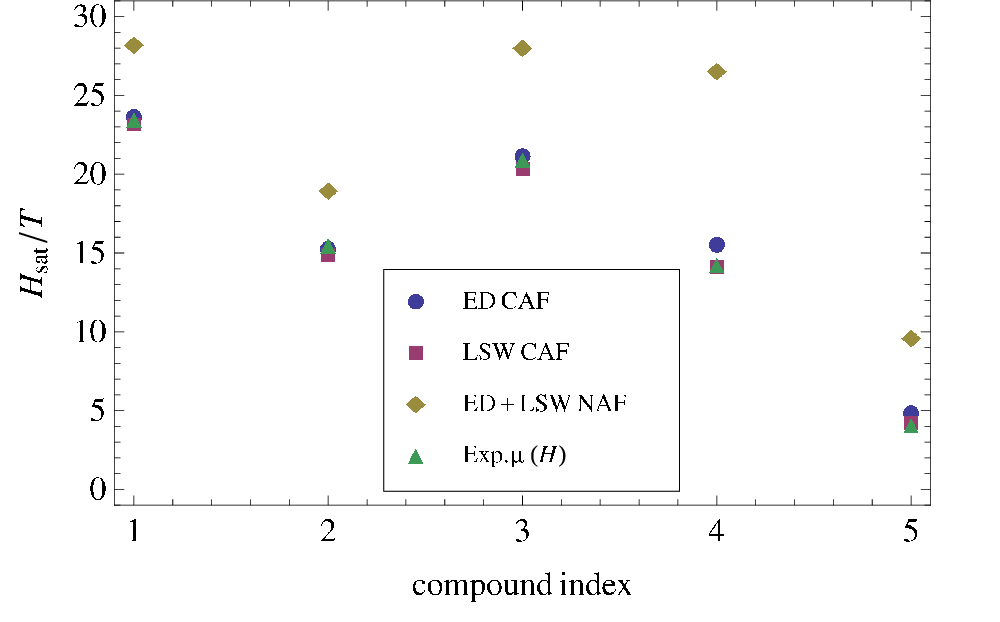}
    \hfill
    \includegraphics[height=0.3\columnwidth]{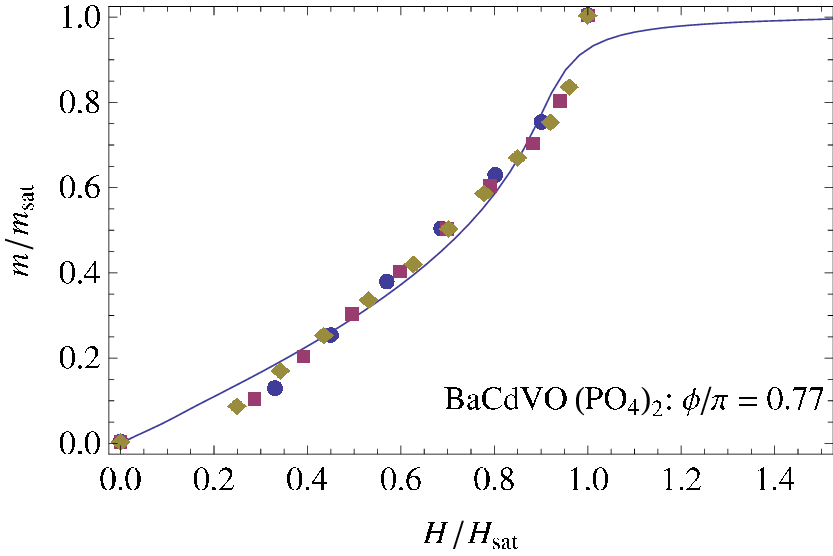}
    \caption{Left: Comparison of saturation fields from ED and
    spin-wave theory with $\theta=\frac{\pi}{4}$, and from high-field
    experiments~\protect\cite{tsirlin:09a}.  The compounds are
    (1)~PbZnVO(PO$_{4}$)$_{2}$, (2)~Na$_{1.5}$VOPO$_{4}$F$_{0.5}$,
    (3)~Pb$_{2}$VO(PO$_{4}$)$_{2}$, (4)~SrZnVO(PO$_{4}$)$_{2}$, and
    (5)~BaCdVO(PO$_{4}$)$_{2}$.  The agreement with the saturation
    fields for the columnar phase gives a direct proof that all
    compounds have CAF order.  Right: Magnetization for
    BaCdVO(PO$_{4}$)$_{2}$~\protect\cite{nath:08}.  Solid line denotes
    experimental data and filled symbols the data from $T=0$ Lanczos
    calculations for different cluster sizes.}
    \label{fig:hsat}
\end{figure}
The left-hand side of Fig.~\ref{fig:hsat} shows a comparison of the
saturation fields determined by LSW and ED for the columnar (CAF) and
Néel (NAF) antiferromagnetic phases with the experimental values
determined from high-field measurements~\cite{tsirlin:09a}.  The
predicted theoretical values are based on fits of our FTLM data and of
a high-temperature series
expansion~\cite{kaul:04,nath:08,tsirlin:09a,rosner:03} to the
temperature dependences of the low-field susceptibilities.  The
experiments agree surprisingly well with the predicted CAF values,
demonstrating that all compounds order in a columnar magnetic
structure at low temperatures.

On the right-hand side of Fig.~\ref{fig:hsat}, the field dependence of
the magnetization for BaCdVO(PO$_{4}$)$_{2}$ is displayed, together
with zero-temperature data from our Lanczos calculations for different
cluster sizes using a Bonner-Fisher
construction~\cite{schmidt:07b,bonner:64}.  Given the small size of
the clusters involved, the agreement is well, apart from low fields,
where finite-size effects are most prominent.

\begin{figure}
    \centering
    \includegraphics[height=0.25\columnwidth]{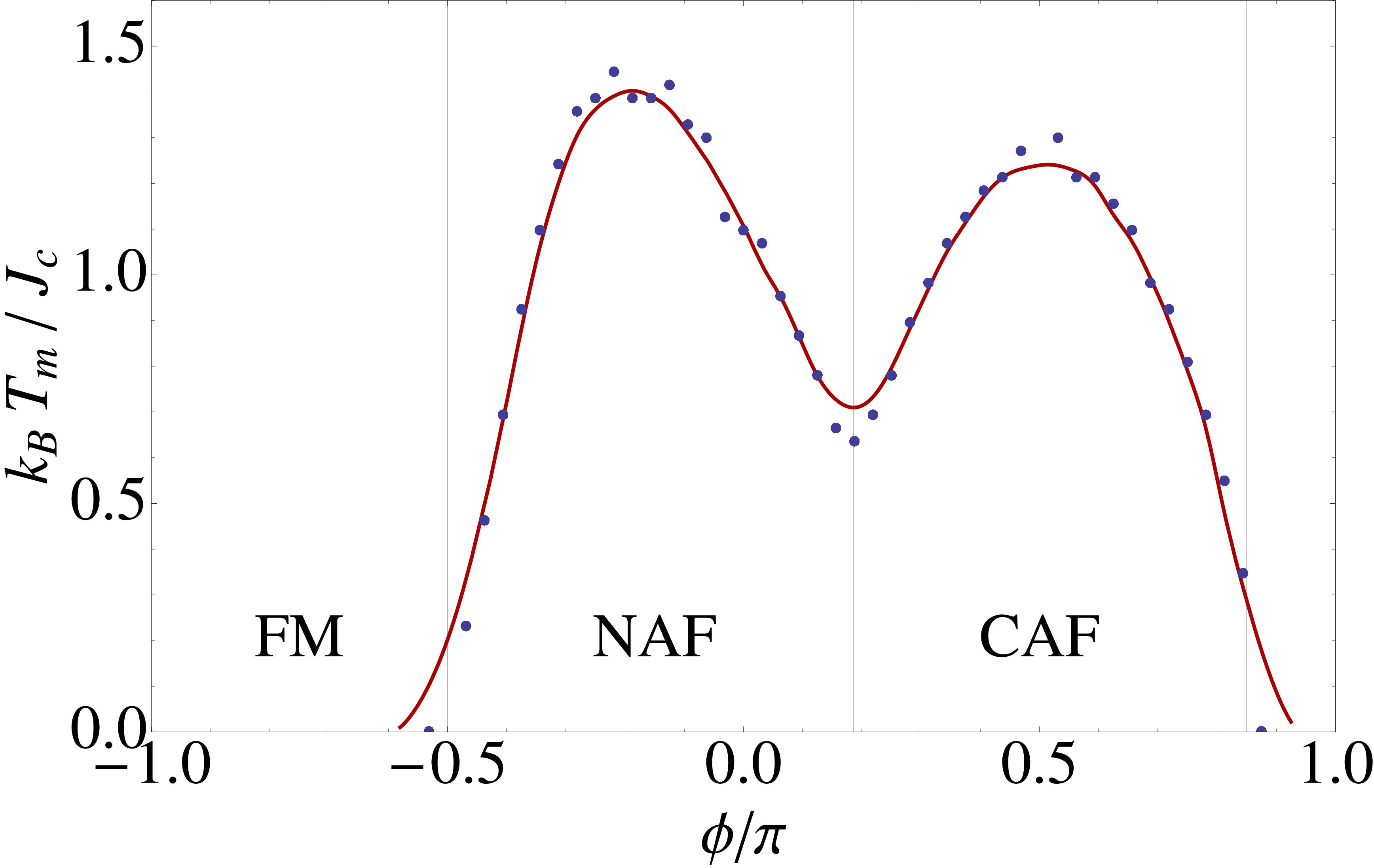}        
    \hspace{0.07\columnwidth}
    \includegraphics[height=0.25\columnwidth]{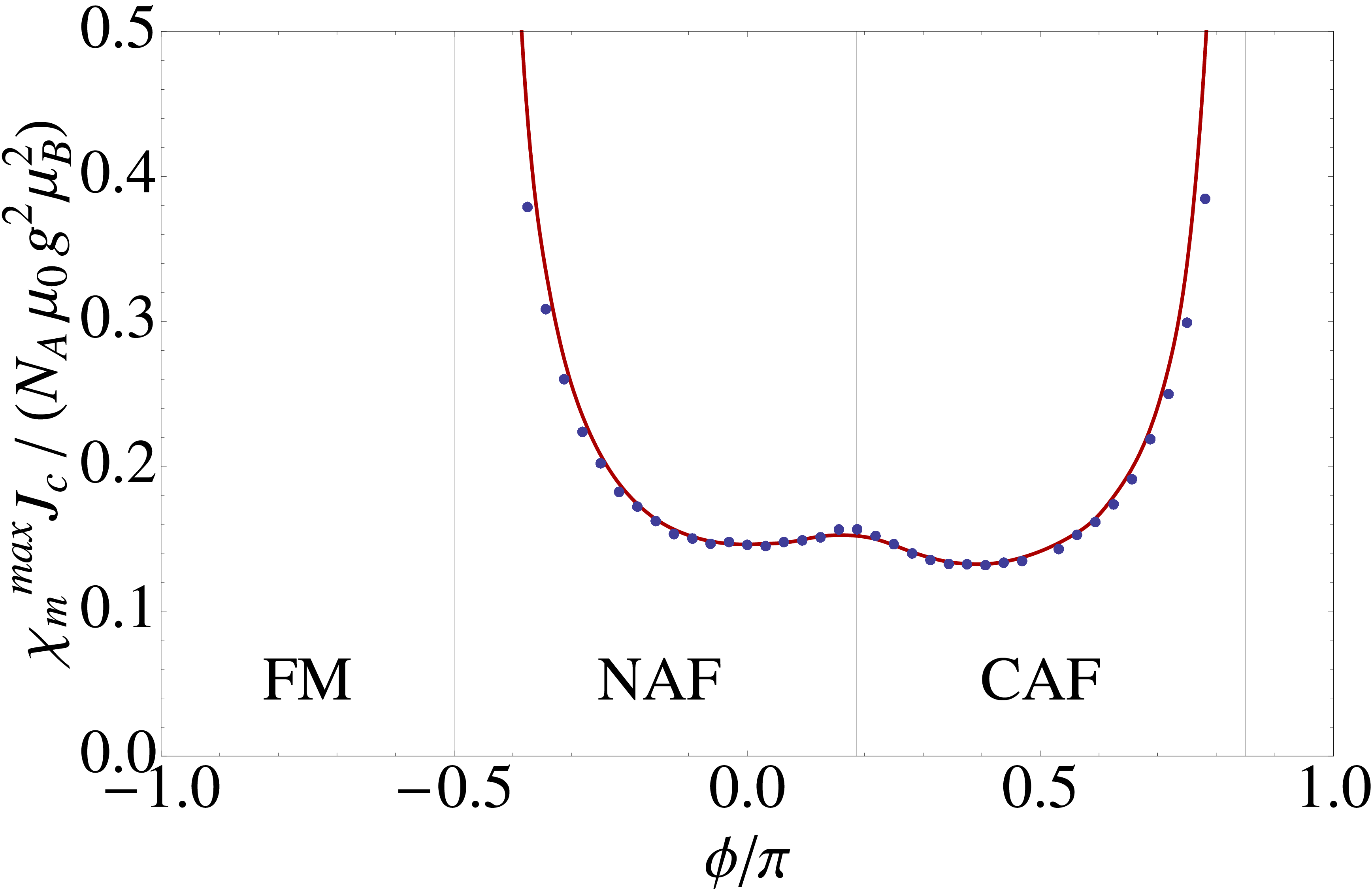}
    \includegraphics[height=0.25\columnwidth]{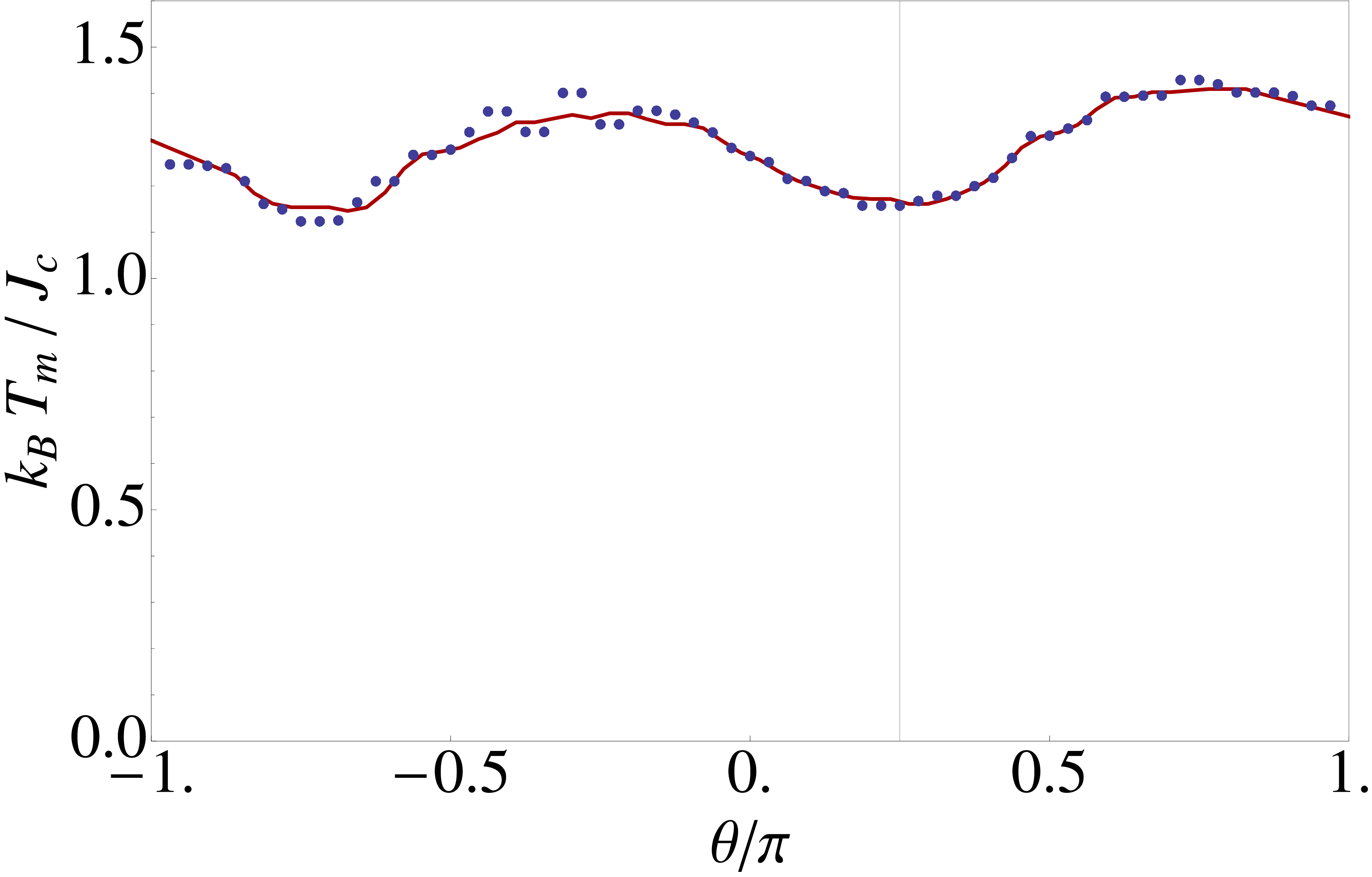}
    \hspace{0.07\columnwidth}
    \includegraphics[height=0.25\columnwidth]{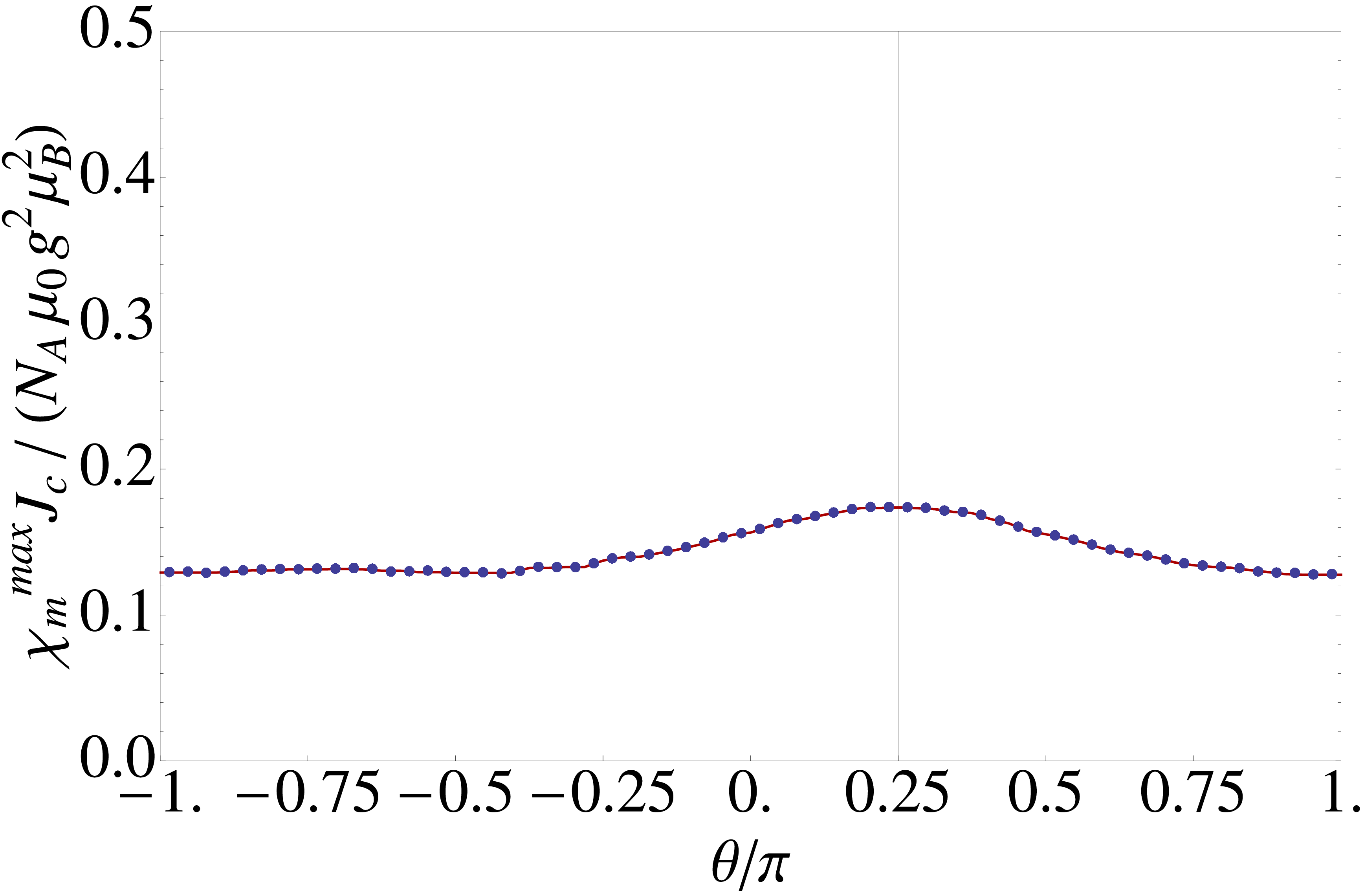}
    \caption{FTLM results for the magnetic susceptibility $\chi(T)$.
    The left-hand figures show the position, the right-hand ones the
    temperature where $\chi(T)$ reaches its maximum.  Top: Dependency
    on the frustration angle $\phi$.  The vertical lines distinguish
    the different classical phases, FM, NAF and CAF. Bottom:
    Dependency on the anisotropy parameter $\theta$ at fixed
    frustration angle $\phi/\pi = 0.625$.  The vertical line shows the
    isotropic case, $\theta=\pi/4$.}
    \label{fig:constphi}
\end{figure}
\section{Extension to the orthorhombic (rectangular) case}

Up to here, we discussed the layered vanadium phosphates in the
context of the square-lattice Heisenberg model.  However, their
crystal structure corresponds to a lattice with lower
symmetry~\cite{tsirlin:09}.  We therefore investigate the impact of an
additional anisotropy of the nearest-neighbor interactions in the ab
plane, characterized by the angle $\theta$ defined in
Eq.~\ref{eqn:params}.

\begin{figure}
    \centering
    \includegraphics[height=0.37\columnwidth]{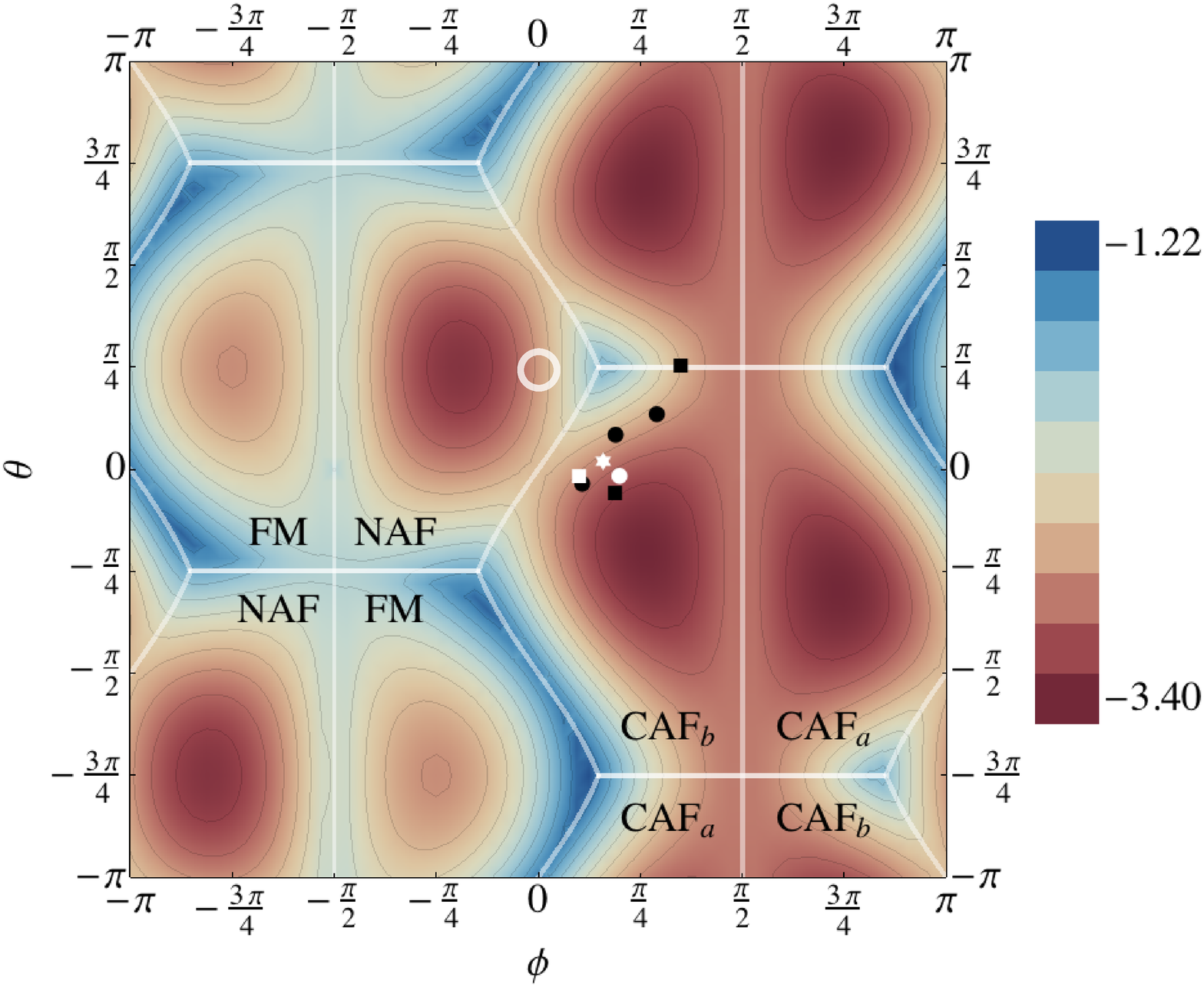}
    \includegraphics[height=0.37\columnwidth]{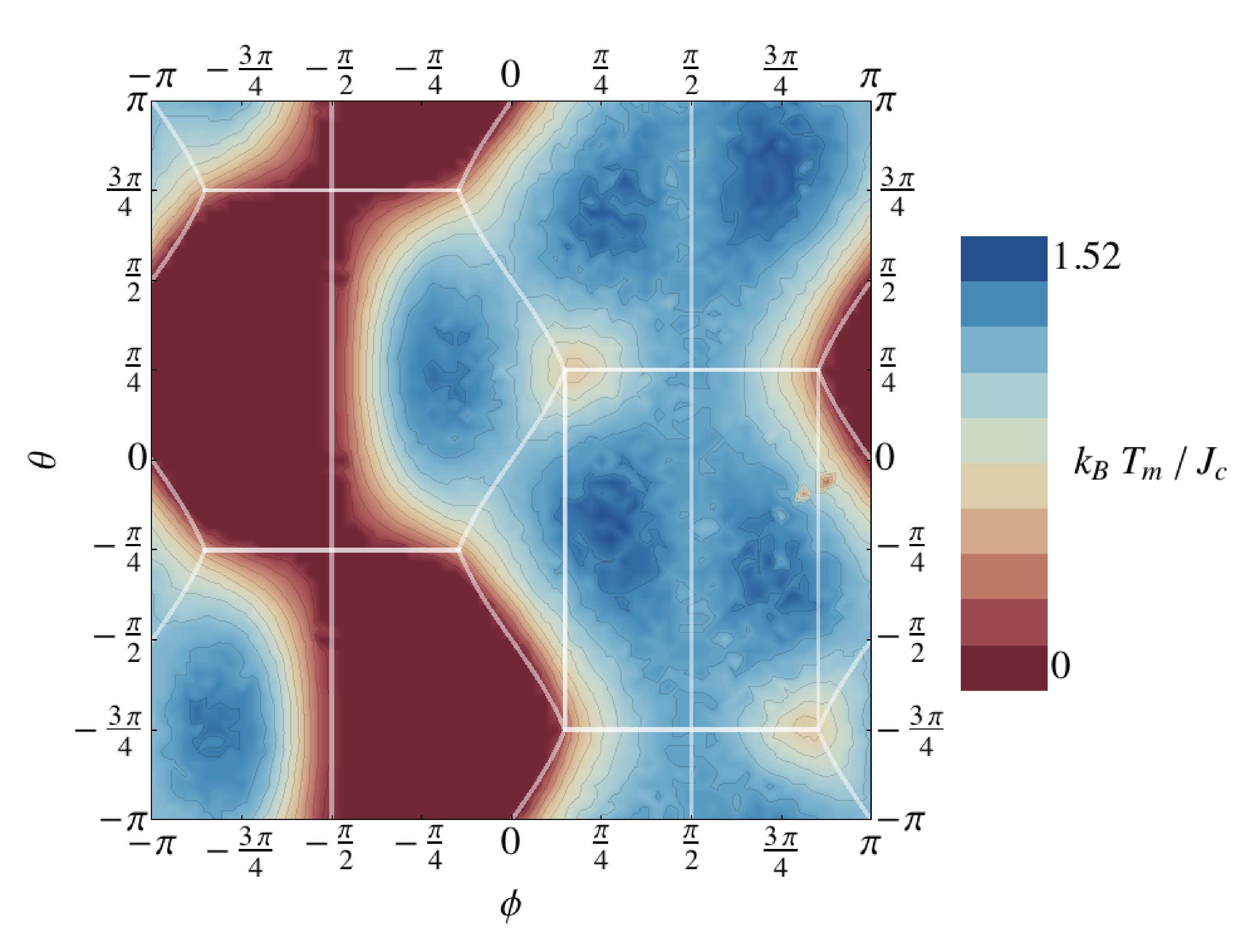}
    \caption{Left: Contour plot of ground state energy as 
    function of anisotropy ($\theta$) and frustration ($\phi$).
    Open circle designates usual NAF ($\theta =\frac{\pi}{4},\phi=0$). Values in table 1
    are represented by black (exp.) and white (theory) symbols.
    The white lines show the boundaries between the four classical
    phases, CAF$_{a,b}$, NAF and FM. Right: Contour plot for the
    position of the maximum susceptibility.}
    \label{fig:gs-chipos}
\end{figure}

In Fig.~\ref{fig:constphi}, FTLM results for a tile of size 20 are
shown.  The position (left) and the value (right) of the broad maximum
of the magnetic susceptibility $\chi(T)$ are plotted as a function of
the frustration ($\phi$) and the anisotropy ($\theta$) angles.  The
top curves correspond to the isotropic case ($J_{1a} = J_{1b}$,
$\theta=\frac{\pi}{4}$), while the bottom ones show the effect of the
anisotropy parameter $\theta$ for constant $\phi/\pi=0.625$ (CAF
regime).  The change both in value and temperature is small.
Therefore introducing an anisotropy within the columnar phase has
comparatively little effect on the temperature dependence of
$\chi(T)$.  For further clarification, a contour plot of the ground
state energy (right) of the Hamiltonian, Eq.~\ref{eqn:ham} and the
temperature of the maximum of $\chi(T)$ (left) as a function of $\phi$
and $\theta$ are shown in Fig.~\ref{fig:gs-chipos}.  The model has
four classical phases, one FM, one Néel AF, and two columnar AF phases
along the crystallographic $a$ and $b$ directions.  Inside the AF
regions, the parameter dependence is weak.  This explains the validity
of the square-lattice Heisenberg model in describing the experimental
results on the thermodynamics of the compounds, which have lower than
tetragonal symmetry.

It has been proposed that magnetic frustration is a key feature of the
magnetic properties of ferropnictides.  However, several different
exchange models are
discussed~\cite{mcqueeney:08,diallo:09,ewings:08,zhao:09}, in
particular models with a spatial anisotropy as described here.  As an
example, Table~\ref{table:xchng} shows the experimental and
theoretical~\cite{han:09} values for moments and exchange constants in
some iron pnictide parent compounds, which are all in the CAF regime.
We conclude that the anisotropy of $J_{1a,b}$ stabilizes the CAF
phase.  In particular the values with $J_{1b}\simeq 0$ and
$J_{1a}/2J_2\simeq 1$ ($\theta \simeq 0, \phi \simeq 0.15\pi$)
correspond to the stable CAF region.  It is obvious from
Fig.~\ref{fig:gs-chipos} (left) that these values are quite distant to
the strongly frustrated point ($\theta = 0.25\pi, \phi = 0.15\pi$)
where the CAF$_{a,b}$ and NAF phases meet.  The moment reduction by
quantum fluctuations for the former values is comparable to that of
the simple unfrustrated NAF (open circle).

\begin{table} 
    \centering
    \small 
    \begin{tabular}{*{9}{l}}
	System&Ref.&$S$&$SJ_{1a}$&$SJ_{1b}$&$SJ_2$&$SJ_c$&$\phi/\pi$&$\theta/\pi$\\ 
	\hline
	CaFe$_2$As$_2$ & \cite{mcqueeney:08} &-& 41    & 10   & 21    & 36   & 0.19 & 0.08 \\
	CaFe$_2$As$_2$ & \cite{diallo:09}    & 0.4& 24-37   & 7-20   & 28-34   & 33-45   & 0.29 & 0.13 \\ 
	CaFe$_2$As$_2$ & \cite{zhao:09}   & 0.22 & 49.9   & -5.7   & 18.9   & 53.7   & 0.11 & -0.04 \\
	BaFe$_2$As$_2$ & \cite{ewings:08} &0.28    & 17.5    & 17.5  & 35    & 39.1 & 0.35 & 0.25 \\
	BaFe$_2$As$_2$ & \cite{ewings:08} &0.54   & 36    & -7   & 18    & 31.6 & 0.19 & -0.06 \\	
	\hline
	CaFe$_2$As$_2$ & \cite{han:09}  &0.75  & 27.4   & -2.1   & 14.5   & 24.3   & 0.20 & -0.02 \\
	BaFe$_2$As$_2$ & \cite{han:09} &0.84   & 36.1   & -2.6  & 12.0    & 38.0 & 0.10 & -0.02 \\
	SrFe$_2$As$_2$ & \cite{han:09}  &0.84  &  35.3    & 2.2  & 13.4    & 28.4 & 0.16 & 0.02 \\
    \end{tabular} 
    \caption{\label{table:xchng}Fe pnictide moment $\mu=2S\mu_B$ and exchange interactions (in meV) from experiment (top)
    and theory (bottom).} 
\end{table} 

\section*{References}

\enlargethispage{.1\baselineskip}
\bibliography{icm_v2}

\providecommand{\newblock}{}
\begin{thebibliography}{10}
\expandafter\ifx\csname url\endcsname\relax
  \def\url#1{{\tt #1}}\fi
\expandafter\ifx\csname urlprefix\endcsname\relax\def\urlprefix{URL }\fi
\providecommand{\eprint}[2][]{\url{#2}}

\bibitem{melzi:01}
Melzi R, Aldrovandi S, Tedoldi F, Carretta P, Millet P and Mila F 2001 {\em
  Physical Review B\/} {\bf 64} 024409

\bibitem{kaul:04}
Kaul E~E, Rosner H, Shannon N, Shpanchenko R~V and Geibel C 2004 {\em J. Magn.
  Magn. Mat.\/} {\bf 272-276} 922

\bibitem{nath:08}
Nath R, Tsirlin A~A, Rosner H and Geibel C 2008 {\em Physical Review B\/} {\bf
  78} 064422

\bibitem{shannon:04}
Shannon N, Schmidt B, Penc K and Thalmeier P 2004 {\em European Physical
  Journal B\/} {\bf 38} 599

\bibitem{schmidt:07b}
Schmidt B, Thalmeier P and Shannon N 2007 {\em Physical Review B\/} {\bf 76}
  125113

\bibitem{tsirlin:09a}
Tsirlin A~A, Schmidt B, Skourski Y, Nath R, Weickert F, Geibel C and Rosner H
  2009 {arXiv}:0907.0391

\bibitem{thalmeier:08}
Thalmeier P, Zhitomirsky M~E, Schmidt B and Shannon N 2008 {\em Physical Review
  B\/} {\bf 77} 104441

\bibitem{rosner:03}
Rosner H, Singh R~R~P, Zheng W~H, Oitmaa J and Pickett W~E 2003 {\em Physical
  Review B\/}  014416

\bibitem{bonner:64}
Bonner J~C and Fisher M~E 1964 {\em Physical Review\/} {\bf 135} A640

\bibitem{tsirlin:09}
Tsirlin A~A and Rosner H 2009 {\em Physical Review B\/} {\bf 79} 214417

\bibitem{mcqueeney:08}
McQueeney R~J {\em et~al.\/} 2008 {\em Physical Review Letters\/} {\bf 101}
  227205

\bibitem{diallo:09}
Diallo S~O {\em et~al.\/} 2009 {\em Physical Review Letters\/} {\bf 102} 187206

\bibitem{ewings:08}
Ewings R~A {\em et~al.\/} 2008 {\em Physical Review B\/} {\bf 78} 220501

\bibitem{zhao:09}
Zhao J {\em et~al.\/} 2009 {arXiv}:0903.2686

\bibitem{han:09}
Han M~J, Yin Q, Pickett W~E and Savrasov S~Y 2009 {\em Physical Review
  Letters\/} {\bf 102} 107003

\end{thebibliography}
\end{document}